\documentclass[twocolumn,twoside,slac_two]{revtex4}
\usepackage{graphicx}
\usepackage{fancyhdr}
\begin{document}

\title{Full Jet Reconstruction in Heavy Ion Collisions: Prospects and Perils}


\author{S. Salur}
\affiliation{Lawrence Berkeley National Laboratory, 1 Cyclotron Road MS-70R0319, Berkeley, CA 94720}

\begin{abstract}
Full jet reconstruction in  heavy ion events has been thought to be difficult due to large multiplicity backgrounds.  A new generation of jet reconstruction algorithms to 
search for new physics in high luminosity p+p collisions at the LHC is developed to precisely measure jets over large backgrounds caused by pile up.  
From simulations it turns out, this new generation of reconstruction algorithms are also applicable in the heavy ion environment. We review the latest results on jet-medium interactions as seen in A+A collisions at RHIC, focusing on the new techniques for full jet reconstruction.

\end{abstract}

\maketitle

\thispagestyle{fancy}


%%%%%%%%%%%%%%%%%%%%%%%%%%%%%%%%%%

\section{Introduction}\label{intro}
 
Jets must be well-defined, measurable from the hadronic final-states and calculable in perturbative QCD (pQCD) from the partonic states \cite{jetsref,seymor}.   The precise measurements of inclusive jet cross sections are performed at many hadronic and leptonic colliders to check in detail the pQCD calculations, to help determine parton distribution functions and to look for new physics. These measurements are found to be in a very good agreement with the different pQCD calculations using various parton distribution functions. The robustness of the theoretical calculations on jet cross sections in $p+\bar{p}$ collisions motivates the use of jets as direct probes of partonic energy loss in the hot QCD matter  generated in ultra-relativistic heavy ion collisions at  RHIC and in near future at the LHC through their interaction and energy loss in the medium (``jet quenching'') \cite{highpt}.   

Until recently, inclusive hadron distributions and di-hadron correlations at high transverse momentum were utilized to measure jet quenching at RHIC indirectly to avoid the complex backgrounds of high multiplicity heavy ion events.   However, these measurements selects jet fragmentation particles that are biased towards the population of jets that has the least interaction with the medium.  Various models with a wide range of parameters are able to describe these measurements however they are leading to only limited constraints upon the underlying physics \cite{bass,armesto,xin}.   Full jet reconstruction in A+A collisions can overcome  theses geometric biases as the energy flow is measured independently of the fragmentation details.  Jet reconstruction at the partonic level with significantly reduced biases enables this study, with qualitatively new observables such as energy flow, jet substructure and fragmentation functions that can be measured in multiple channels (inclusive, di-jets, h-jets and gamma-jets).

The detector upgrades together with the increased beam luminosities of RHIC and data recording capabilities of  PHENIX and STAR, enable jet reconstruction in heavy ion collisions for the first time \cite{salurww}.  In this article we review the prospects and the perils of the new techniques developed for full jet reconstruction in heavy ion collisions at RHIC and the latest results obtained with these new tools. The experimental details of jet reconstruction measurements  utilizing the STAR and PHENIX experiments can be found in \cite{me, ploskonQM, yuiQM, yuiDPF} for the inclusive spectra,  \cite{cainesQM,kapitanQM} for the underlying event, and  \cite{jor,brunaQM} for the accompanying jet fragmentation studies in heavy ion collisions.

\section{Prospects}

Various jet reconstruction algorithms have been developed for both leptonic and hadronic colliders during the last 20 years.   %See  \cite{me,davidE} and the references therein for an overview of jet algorithms in high energy collisions.    %Figure~\ref{fig:dijets} shows an example of an identified di-jet event from a central Au+Au collision, using both the neutral energy from the Barrel Electro-Magnetic Calorimeter and charged particles from the Time Projection Chamber of the STAR experiment. Clear di-jet peaks can be observed over the underlying heavy ion environments.    
  Here we will briefly discuss the cone and sequential recombination algorithms that are used for the STAR analysis.  Further algorithmic details of cone,  sequential recombination and Gaussian filtering can be found in \cite{catchment,jets,kt,ktref,blazey, gauss, gauss1} and the references therein.

%\begin{figure}[here!]
%\begin{center}
%\resizebox{0.48\textwidth}{!}{
%\includegraphics{starjetpict2.eps}}
%\caption[]{A reconstructed di-jet from a central Au+Au event at $\sqrt{s_{NN}}=200$ GeV in the STAR detector \cite{me,jor}. } \label{fig:dijets}
%\end{figure}

\subsection{Jet Reconstruction Algorithms}

The cone algorithm is based on the simple picture that a jet consists of a large amount of hadronic energy in a small angular region. Therefore, the main method for the cone algorithm is to combine particles in $\eta - \phi $ space with their neighbors within a cone of radius R ($\rm R=\sqrt{ \Delta \phi ^{2}+ \Delta \eta^{2} }$). To accommodate higher-order processes and to optimize the search and effectiveness of jet finding, splitting, merging, and iteration steps can be used.

  Unlike the cone algorithm, the sequential recombination algorithms  combine pairs of objects relative to the closeness of their $p_{T}$. Particles are merged into a new cluster via successive pair-wise recombination.  While the lowest $p_{T}$ particles are the starting point for clustering in the $k_{T}$ algorithm, in the anti-$k_{T}$ algorithm, recombination starts with the highest momentum particles.  Due to these different approaches, in the $k_{T}$ algorithm, arbitrarily shaped jets are allowed to follow the energy flow, resulting in less bias on the reconstructed jet shape than with the anti-$k_{T}$ or cone algorithm which are more or less restricted to a circular shape \cite{catchment}.

Motivated by the need for precision jet measurements in the search for new physics in high luminosity p+p collisions at the LHC, a new approach to jet reconstruction and background subtraction was developed  \cite{catchment,salamtalk}. A key feature in this procedure is a new QCD inspired algorithm for separating jets from the large backgrounds due to pile up. As it turns out from simulations,  these improved techniques can also be used in heavy ion environments where the background subtraction is essential for jet measurements.  Sequential recombination algorithms ($\rm k_{T}$, anti-$\rm k_{T}$ and Cambridge/Aachen (CAMB)) encoded in the $FastJet$ suite of programs  \cite{catchment,antikt}, along with a seedless infrared-safe cone algorithm (SISCone) \cite{sis} are utilized to search for jets in p+p \cite{elena} and Au+Au collisions collected by the STAR experiment.  

An alternative seeded cone algorithm  was also explored previously by STAR experiment for Au+Au collisions in order to avoid instabilities in cone-finding due to large heavy ion background \cite{me,jor}.   In addition to recombination and cone algorithms, the Gaussian filtering type of algorithm that simply extracts jets as local maxima in the $\eta-\phi$ space by linearly filtering particles is also used by the PHENIX experiment to extract jets successfully in p+p and Cu+Cu collisions \cite{yuiQM, yuiDPF}.

\section{Perlis}

The complex heavy ion background makes the full jet reconstruction a challenging task. Here we discuss the fundamental assumptions required by the full jet reconstruction in these complex environments and the new biases that can be introduced.  These perils require further investigation of the algorithmic responses of full jet reconstruction in heavy ion collisions by utilizing various reconstruction algorithms and the implications on jet quenching in heavy ion collisions. 

\subsection{Underlying Heavy Ion Background}
 
 A fundamental assumption that the signal and the uniform background are two separable components is required by the full jet reconstruction in heavy ion collisions.  However, this assumption can be violated by the presence of jets and the effect of these jets on the background estimation. For example the initial state radiation, even though expected to be small compared to jet energy, might be different in A+A relative to p+p.  Initial state processes resulting in the enhancement of the multiplicity of the underlying events appears to be distributed uniformly in the simpler p+p case \cite{cainesQM} and therefore it is fully accounted for the estimation of the background under jets \cite{catchment}. However in the Au+Au collisions, ``the p+p correspondent" underlying event may be modified, possibly generating non-uniform structures. These non-uniformities in the background might be even larger  due to the final state processes.  Energy loss of the jet in matter might modify the event shape, resulting in non-uniform structures such as the ridge. Azimuthal and longitudinal anisotropy of heavy ion events will also result into non-uniform backgrounds.  
  Some of these sources of correlated backgrounds can be brought under quantitative control by using different collision systems. On the other hand, other observed effects might help us to understand details of the jet interactions with the heavy ion environment and may give further insight into the structures that are observed in di-hadron correlations and their origins.

\begin{figure*}[top!]
\resizebox{0.95\textwidth}{!}{
\includegraphics{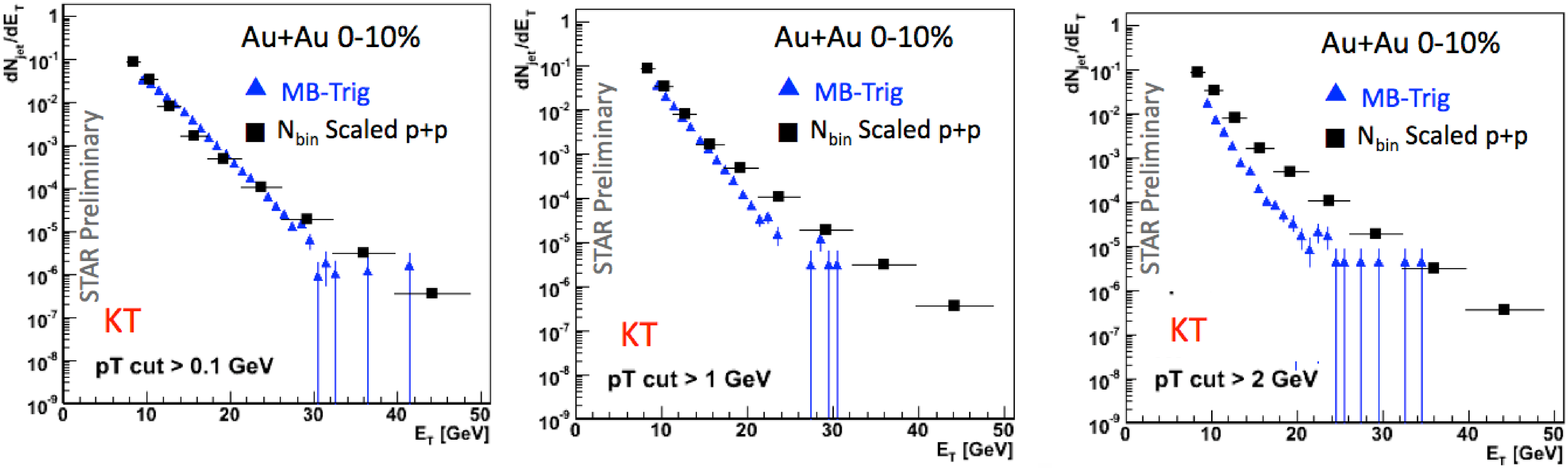}}
 \caption[]{The $\rm p_{T}$ threshold dependent comparison of inclusive jet spectra in heavy ion and p+p collisions.
The filled triangle symbols are from MB-Trig, and filled squares are from $\rm N_{Binary}$ scaled p+p collisions.
In addition to $\rm p_{T}$ cuts shown in the figures, R is set to 0.4. \cite{me}.
 } \label{fig:ptcut}
\end{figure*}

With the assumption that the signal and the background are two separable components, the background correction can be estimated by following three simple steps.  The first step is measuring the jet area for the infrared safe algorithms as encoded in the FastJet Suite of algorithms. An active area of each jet is estimated by filling an event with  many very soft particles and then counting how many are clustered into a given jet. The second step is measuring the diffuse noise  (mean $p_{T}$ per unit area in the remainder of the event) and noise fluctuations.  Event-by-event fluctuations in the background also distort the jet spectrum towards larger  $E_{T}$ due to the steeply falling dependence of the jet production on $E_{T}$. This effect can be corrected  through an unfolding procedure (i.e., deconvolution).   The final step is correcting the jet energy by deconvolution of signal from the complex heavy ion background, using parameters that are extracted from measurable quantities of area, diffuse noise and noise fluctuations.

%Elliptic flow and the longitudinal non-uniformity of background can be measured with a good accuracy in d+Au events and the results then can be applied to estimate the effect in Au+Au measurements.  The possible structures due to the initial and final state effects of heavy ion environment can  be also further studied with the jet reconstruction by looking into the jet profile. 
% These caveats further need to be investigated with different jet algorithms and their various responses to the heavy ion background for a full understanding of the algorithmic full jet reconstruction in heavy ion collisions and the heavy ion environment resulting in jet quenching. 

\subsection{Biases}

The ultimate goal of full jet reconstruction is to investigate the jet quenching in heavy ion collisions at the partonic level, without any ambiguities being introduced by hadronization and geometric biases of the inclusive spectrum and di-hadron measurements.  However,  it is possible that  new biases can be introduced when reconstructing jets. For example, all jet algorithms have various parameters for searching and defining jets, and the effects of varying these parameters need to be explored  in detail for a full understanding of jet reconstruction.  

A bias will be introduced while trying to reduce the effect of the background fluctuations in heavy ion collisions with the threshold cuts on the track momenta and calorimeter tower energies ($p_{T}^{cut}$). Figure~\ref{fig:ptcut} shows the comparison of the jet spectra reconstructed by the $k_{T}$ algorithm with a variation of $p_{T}$ threshold cuts in Au+Au and the $\rm N_{Binary}$ scaled p+p collisions \cite{me}. 
For these results, energy resolution of the detectors and the underlying heavy ion background fluctuations were corrected with multiplicative factors of the jet spectrum estimated  by utilizing  Monte-Carlo model studies based on Pythia 8.107 \cite{pythia}.  While the agreement between Au+Au and $\rm N_{Binary}$ scaled p+p  jet measurements is good for the lowest value of the $p_{T}$ cut, it is also seen to be poorer with the larger
$p_{T}$ threshold cut. This suggests that the threshold cuts introduce biases which are not fully corrected by the procedures that use fragmentation models that are developed for $\rm e^{+} + e^{-}$ and p+p collisions.

To enhance the recorded rate of high $p_{T}$ particles and jets, events above some threshold in the electromagnetic calorimeter  are collected. (This threshold is 5.4 GeV for the STAR experiment during data taking in years 2006 and 2007.)   This is very similar to the case of jets that are reconstructed with seeded infrared unsafe algorithms.  
The online calorimeter tower triggers introduce a strong bias of reconstructed jets that are fragmenting hard in comparison to the jets that are reconstructed without a seed. 
%New observables such as intra-jet energy distributions, jet-jet and hadron-jet correlation studies can be used to further elucidate the effect of these biases. 

The resolution parameter or cone size, which restricts the area of the jet and thereby the amount of energy flow, can be a harder parameter to calculate hence interpret in heavy ion collisions than in p+p collisions.  If the jets are broader  in the heavy ion environment, the same resolution parameter might not be sufficient to recover the same fraction of jet energy in comparison to p+p jets.  This bias needs to be investigated by varying the resolution parameter and by looking into the jet profile of these jet definitions.

%However this method was known to be approximate, as the full processing of an input Monte-Carlo spectrum is predicted to be modified due to quenching. The observed spectrum also involves additional convolution with the energy resolution plus losses due to acceptance and efficiency. 

\section{Results}

\begin{figure}[h!]
%\begin{center}

\centering

\resizebox{0.45\textwidth}{!}{
\includegraphics{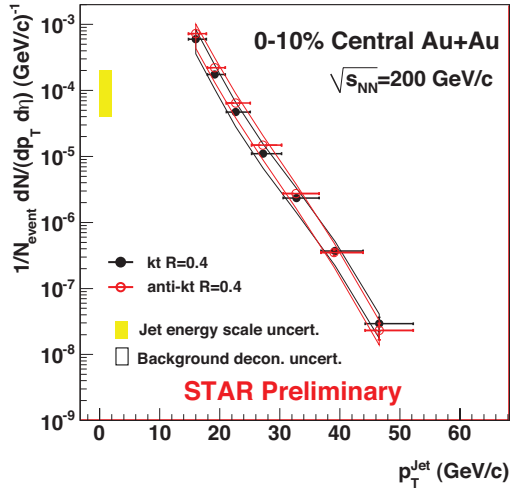}}

 \caption[]{Inclusive jet yield per event vs transverse jet energy for the central Au+Au collisions obtained by the sequential recombination ($\rm k_{T}$ and anti-$\rm k_{T}$) algorithms  \cite{ploskonQM}. } \label{fig:kt}
\end{figure}

The Figure~\ref{fig:kt} shows the comparison of the inclusive jet spectra reconstructed by $\rm k_{T}$ and anti-$\rm k_{T}$ sequential recombination algorithms for central Au+Au collisions collected by the STAR experiment \cite{ploskonQM}.  Systematic uncertainties due to the unfolding procedure  are shown as the envelopes in red and black and the jet energy resolution as the yellow bar.  Jet spectra are consistent between the two  sequential reconstruction algorithms extending to 50 GeV kinematic reach. Corrected jet spectra reconstructed with a Gaussian filtering algorithm for various Cu+Cu centralities collected by the PHENIX experiment within their restricted experimental acceptance were also presented in this meeting \cite{yuiQM, yuiDPF}. The same algorithms that are used for heavy ions are also used to reconstruct jets in p+p collisions. The jet spectra reconstructed in p+p collisions collected by PHENIX and STAR experiments (with sigma of 0.3 for gaussian filtering and the resolution parameter of 0.4 for the sequential recombination algorithm) all agree well with the previously published RHIC results using a cone algorithm with split merge steps (cone radius of 0.4) \cite{ploskonQM,yuiQM,starpp,nlo}.

 %Based on scaling of the $p+p$ spectrum shown in Figure 1, the Au+Au dataset from the 2007 RHIC run is expected to contain statistically significant jet yield beyond 50 GeV. However, measuring jets above the complex heavy ion background is a challenging task. 

\begin{figure}[h!]
%\begin{center}

\centering
\resizebox{0.45\textwidth}{!}{
\includegraphics{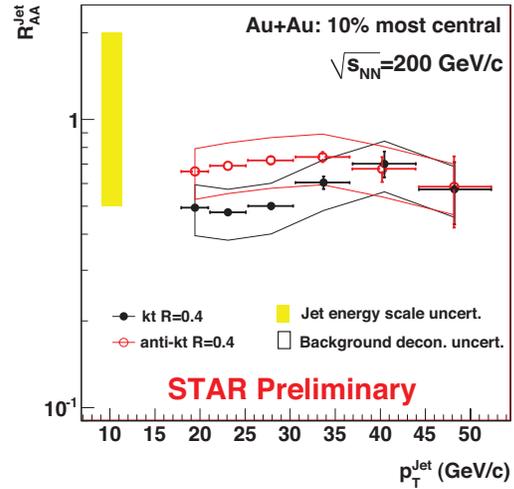}}

 \caption[]{Momentum dependence of the nuclear modification factor of jet spectra reconstructed with $ \rm k_{T}$ and anti-$\rm k_{T}$ algorithms (0-10\% most central Au+Au divided by $N_{Binary}$ scaled p+p collisions) \cite{ploskonQM}.  
 
   } \label{fig:raa}
\end{figure}

 The nuclear modification factor ($\rm R_{AA}$) for the reconstructed jet spectra with the resolution parameter of 0.4 from $\rm k_{T}$  and anti-$\rm k_{T}$ can be found in Figure~\ref{fig:raa}.  The envelopes shown represent the one sigma uncertainty of the deconvolution of the heavy ion background. The total systematic uncertainty on the jet energy scale is around 50\%, shown as the yellow bar.  In the case of full jet reconstruction, $\rm N_{Binary}$ scaling as calculated by a Glauber model \cite{glauber} is expected if the reconstruction is unbiased, i.e. if the jet energy is recovered fully independent of the fragmentation details, even in the presence of strong jet quenching. This scaling is analogous to the cross section scaling of high $p_{T}$ direct photon production in heavy ion collisions observed by the PHENIX experiment \cite{phenix}.   A large fraction of jets are reconstructed  when using both $\rm k_{T}$ and anti-$\rm k_{T}$ sequential recombination algorithms with a resolution parameter of 0.4. Momentum dependence of the nuclear modification factor is also different than the observed suppression of the $\pi$ meson $\rm R_{AA}$. However even though there are large systematic uncertainties, a hint of a suppression of jet $\rm R_{AA}$ above 30 GeV can be observed. This implies that an unbiased jet reconstruction is not reached fully.  It is expected that for a smaller resolution parameter, this suppression should reach to single particle suppression at large momentum. For the case of R=0.2, further suppression is observed to grow to larger values supporting this expectation \cite{ploskonQM}.

\begin{figure}[h!]
%\begin{center}

\resizebox{0.49\textwidth}{!}{
\includegraphics{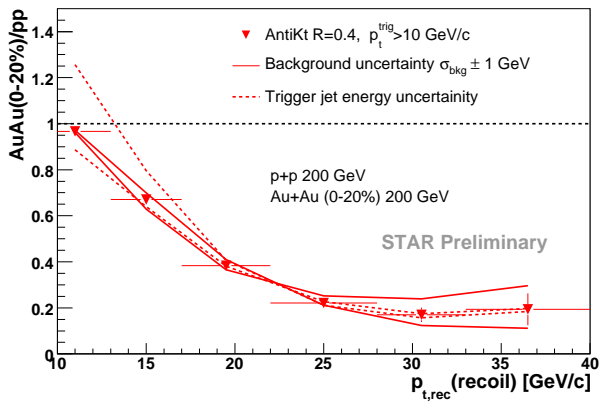}}

 \caption[] {The ratio of the spectra of the recoil recoil jets  in 0-20\% central High Tower triggered Au+Au events to High Tower triggered p+p collisions \cite{brunaQM}.  
   } \label{fig:dijetratio}
\end{figure}

The ratio of the spectra from the recoil jets of the di-jet coincidence measurements in 0-20\% central Au+Au to p+p collisions is presented in Figure~\ref{fig:dijetratio} \cite{brunaQM}.  The recoil jets are selected when the triggered jets have $p_{T}$ greater than 10 GeV and 20 GeV as presented as the triangles and the open circles. The recoil jet spectra are normalized to the number of trigger jets.  Background and trigger jet energy uncertainties are shown with the solid and dashed histograms.  While a large fraction of inclusive jets are reconstructed yielding a much smaller suppression as seen in Figure~\ref{fig:raa}, with these trigger selections biased population of recoil jets are measured resulting into a suppression that is much more comparable to the measurement of $\pi$ meson $\rm R_{AA}$. It is possible to introduce geometric biases that are similar to the observed di-hadron measurements when reconstructed jets are selected with a given criteria.

Medium effects at parton splitting can also be studied with fragmentation functions. Fragmentation function measurements are extremely challenging due to the uncertainty in the statistical subtraction of the background particles and to potential biases in the reconstruction discussed earlier.  Nevertheless, the Figure~\ref{fig:raa} is a first attempt of the $z$ ($z=p_{T}^{hadron}/p_{T}^{Jet}$) dependence of the ratio of the fragmentation functions from the recoil jets of the di-jet coincidence measurements in 0-20\% central Au+Au to p+p collisions \cite{brunaQM}. See the inset of the right panel in Figure~\ref{fig:raa} for the applied selection cuts for the trigger and the recoil jets reconstructed with the anti-$k_{T}$ algorithm. Various uncorrelated systematic uncertainties are also included as the solid and the dotted histograms.  Within the given systematic uncertainties, the ratio of the fragmentation functions of Au+Au to p+p show no significant suppression. The lack of modification in fragmentation functions confirms that the di-jets that are reconstructed in this analysis are biased towards a sample of unquenched jets that are only coming from the surface as also observed as the suppression in Figure~\ref{fig:dijetratio}.  It is also possible that due to the broadening of the jets and the requirement of the limited resolution parameter only part of the jet energy is recovered. This results in an insensitivity to expected softening when measuring fragmentation functions statistically.

\begin{figure}[h!]
%\begin{center}

\resizebox{0.52\textwidth}{!}{
\includegraphics{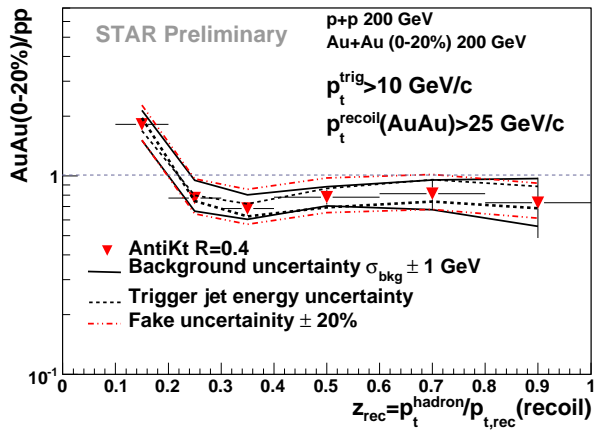}}

 \caption[]{$z$ dependence of the ratio of the fragmentation functions for recoil jets (normalized to the number of recoil jets) 
measured in 0-20\% central High Tower triggered Au+Au events to High Tower triggered p+p collisions \cite{brunaQM}.  The systematic uncertainties in the estimation of the width of the Gaussian parameterization of the background fluctuations are presented as solid envelopes. 
 
   } \label{fig:ff}
\end{figure}

%or just due to shear luck the effect of quenching is compensated with the un-recovered energy due to broadening of jets and limited resolution parameter. 

 \section{Conclusions}

 Full jet reconstruction expands the kinematic reach to much larger values.  These large momenta reaching 50 GeV can be studied in heavy ion collisions for the first time. New physics effects should be considered when interpreting the results at large momentum. A possible effect might be the momentum dependence of the relative contributions of quark and gluon sub-processes to inclusive jet production. In elementary p+p collisions these contributions vary with respect to the jet momentum \cite{vogelsang}.  The relative contributions might be even different in a heavy ion environment when a quark gluon plasma is produced, affecting the expected shape of the jet spectra and therefore of the nuclear modification factors. Another possibility is that at large momentum fraction $x$, initial state effects (such as the EMC effect which is the deviation between structure functions of heavy ions to light ions) are observed to be as large as 15\% \cite{emc}.  There might be other effects like the EMC effect playing a major role in the relative suppression or enhancement of nuclear modification factors at large momentum.  %To further investigate the quenching dependence on parton species in heavy ion collisions and the applicability of the jet reconstruction methods new observebles need to be measured. 

The new Monte-Carlo based simulations of  jet quenching in medium such as Jewel \cite{jewel}, Q-Pythia \cite{qpythia} and YaJEM \cite{yajem} and complementary analytic  calculations \cite{vitev,vitev2, vitevDPF, borghini}  recently became  available to pursue a quantitative analysis of jet quenching as observed in heavy ion collisions.  However there are many uncertainties (e.g., how hadronization is treated) in the predictions  of these models and calculations.  To confront the calculations with data, new robust QCD jet observables that are unaffected by the $p_{T}$ cuts and hadronization need to be explored experimentally. For example the  subjet observable is infrared safe and insensitive to hadronization and may be used to study the jet quenching \cite{jewel}.  %Due to the medium induced radiation more coarser jet structures are observed in Jewel.

Unbiased reconstruction of jets in central heavy ion collisions at RHIC energies would be a breakthrough to investigate the properties of the hot QCD matter. 
The studies shown here indicate that reconstruction of jets with a uniquely large kinematic limit may indeed be possible in heavy ion events. However jet reconstruction in heavy ion collisions is not yet free of biases.  Biases are introduced due to selection of particles such as $p_{T}$ cuts to reduce the fluctuations of heavy ion background, requirement of algorithmic parameters such as cone size or resolution parameter, and the collection of events with thresholds to enhance the jet rates. To assess fully the systematic uncertainties of jet measurements these effects must be investigated further. % While with full jet reconstruction in heavy ion collisions, a much more reduced geometric bias is reached, other biases such as data taking, selection of particles such as $p_{T}$ cuts and resolution parameters  must be investigated further in order to assess fully the systematic uncertainties of this measurement. 
These results motivate us to look further into multiple channels for consistency checks (inclusive, di-jets, h-jets, gamma-jets to measure qualitatively new observables: energy flow, jet substructure, fragmentation function \cite{vitev,vitev2}.)

%A copious production of very energetic jets, well above the heavy ion background, is predicted to occur at the LHC. The large kinematic reach of high luminosity running at RHIC and at the LHC may provide a sufficient lever-arm to map out the QCD evolution  of jet quenching \cite{solan}.  The comparison of  full jet measurements in the different physical systems generated at RHIC and the LHC will provide unique and crucial insights into our understanding of jet quenching and the nature of hot QCD matter.

 %%However, spectrum corrections are currently based on model calculations using PYTHIA fragmentation \cite{PYTHIA}. This aspect, together with the spectrum variations due to cuts and reconstruction algorithms, must be investigated further in order to assess the systematic uncertainties of this measurement. 

%Further, we utilize the reconstructed jets and study the jet shapes to test the underlying QCD theory \cite{vitev,vitev2}.  The results from the intra-jet energy distributions, jet-jet and hadron-jet correlation studies will be available in the coming months and will enable us to study the medium properties produced at RHIC. Also 

\section*{Acknowledgments}
The author wishes to thank Yui-Shi Lai and Ivan Vitev for the valuable discussions on the topic of this proceeding during the meeting and the organizers for the invitation and for the fruitful meeting.


\end{document}